\documentclass[twocolumn,prl]{revtex4}
\usepackage{graphicx}
\begin{document}

\title{Spin freezing and magnetic inhomogeneities
in bilayer manganites}

\author{A.\ I.\ Coldea$^1$, S.\ J.\ Blundell$^1$, C.\ A.\ Steer$^1$,
J. F. Mitchell$^2$, F. L. Pratt$^3$}
\address{$^1$Clarendon Laboratory, University of Oxford, Parks
Road, Oxford OX1 3PU, UK}
\address{$^2$Materials Science Division,
 Argonne National Laboratory, Argonne, Illinois 60439, USA}
\address{$^3$Rutherford Appleton Laboratory, Chilton, Didcot, OX11 0QX, UK}

\begin{abstract}
We have performed a muon spin rotation study on polycrystalline samples of
electron-doped layered manganites,
La$_{2-2x}$Sr$_{1+2x}$Mn$_2$O$_7$ ($0.4 \leq x<1$),
in order to investigate
the local magnetic structure
and spin dynamics.
Our results
provide evidence for phase separation into A-type
antiferromagnetic and charge-ordered phases for
$x$=0.52 and spin freezing
at low temperatures ($T<100$~K) for $0.52 \leq x<0.75$.
A new phase diagram which includes this spin
freezing region is proposed.

\end{abstract}


\maketitle

The phenomenon of
phase separation
is crucial for
determining the
intrinsic properties of many
magnetic oxides \cite{Dagotto2001}.
 The origin of the colossal magnetoresistance (CMR) in manganese
perovskite structures is often linked
to the coexistence of ferromagnetic (FM) metallic
and antiferromagnetic charge
ordered (CO) insulating domains
or to the formation of nanoscale magnetic clusters
\cite{Uehara1999,Teresa1997N}.
One technique which provides information on the local
magnetic order is muon spin rotation/relaxation ($\mu$SR).
In underdoped cuprates, $\mu$SR has identified
the coexistence of superconducting and
antiferromagnetic phases as well as spin freezing believed
to be related to the generation of stripes \cite{Niedermayer1998}.

In this letter we provide direct evidence for spin-freezing and
phase separation in electron-doped layered manganite structures
using $\mu$SR. We investigate the local magnetic structure and
spin dynamics across a series of magnetic structures of
La$_{2-2x}$Sr$_{1+2x}$Mn$_2$O$_7$ with $0.4 \leq x<1$. For
compounds in the Mn$^{4+}$-rich half of the phase diagram
($0.5<x<1$) neutron diffraction studies \cite{Ling2000} reveal a
progression of antiferromagnetic (AFM) insulating phases: from
A-type ($0.5 \leq x< 0.66$) to C/C$^*$-type ($0.75\leq x< 0.9$)
and to G-type ($0.9<x \leq 1.0$). Furthermore, no long-range
magnetic order (NLRO) has been observed between the A-type AFM and
C/C$^*$-type AFM regions ($x$=0.66-0.74) \cite{Ling2000}. Our
experiments measure the temperature dependence of the order
parameter in the samples with long-range order (LRO) and show how this is
affected by the interplay between magnetic and charge order.
Moreover, by cooling a sample in the NLRO region, we are able to
follow the progressive slowing down of magnetic fluctuations and
demonstrate the development of short-range order. Based on our
results we are able to propose a new phase diagram of
La$_{2-2x}$Sr$_{1+2x}$Mn$_2$O$_7$ ($0.5 \leq x<1$),
supplementing the information already
obtained from neutron diffraction \cite{Ling2000}.

Zero-field $\mu$SR is especially suited for study of short-range magnetic
correlations since the positive muons are a sensitive local probe.
Because the local magnetic fields at muon sites result primarily
from dipolar interactions which decay very quickly with increasing
distance ($\sim 1/r^3$) the effective range investigated with muons is
$\sim 20~$\AA. In a system having magnetically ordered and
disordered regions the $\mu$SR data is composed of two different
signals corresponding to different environments, with signal amplitudes
roughly proportional to their volume fraction.
In the case of a magnetically ordered system
our data are fitted using a relaxation function, $G_z(t)$, of the
form
\begin{equation}
\small
 G_z(t)=A_{\rm rlx}\exp(-t/T_{1})
 +A_{\rm osc}\cos(2\pi\nu_{\mu}t)\exp(-t/T_{2})
  \label{ch3:Gztotal}
  \end{equation}
where the first term is the longitudinal (spin-lattice) relaxation, the second
term is the muon precession at frequency
$\nu_{\mu}$=$\gamma_{\mu}{B_{\mu}}/2\pi$ in the local internal
field, $B_{\mu}$, in the ordered state, $1/T_2$ is the
transverse relaxation rate and
 $\gamma_{\mu}/2\pi$=135.5~MHz/T is the gyromagnetic ratio.
The longitudinal relaxation rate is defined by
$1/T_1 \propto  (\gamma_{\mu} \Delta B_{\mu})^2 \tau$
for rapid fluctuations,
where  $\Delta B_{\mu}$
is the amplitude of the fluctuating field and $\tau$ is the Mn-ion
correlation time. Near a magnetic phase transition $\tau$
 (and hence $1/T_1$) increases with cooling due to the critical
slowing down.
 For polycrystalline samples,
where the local magnetization points along the muon spin direction
with probability 1/3, the corresponding amplitudes are
$A_{\rm rlx}= G_z$(0)/3 and $A_{\rm osc}= 2G_z$(0)/3.
Eq.~\ref{ch3:Gztotal}
can be generalized in a straightforward manner if there are
multiple frequencies in the data corresponding to magnetically
inequivalent sites. Polycrystalline samples ($\sim0.5$~g) were
prepared as described in Ref. \cite{Millburn1999}.
Zero-field $\mu$SR data for 10 compositions
were taken on the EMU beamline at the ISIS
facility (Rutherford Appleton Laboratory, UK) and for 5 compositions
on the GPS
spectrometer at the PSI muon facility (Paul Scherrer Institute,
Switzerland).

\begin{figure}[htb]
\centering
  \includegraphics[width=6cm]{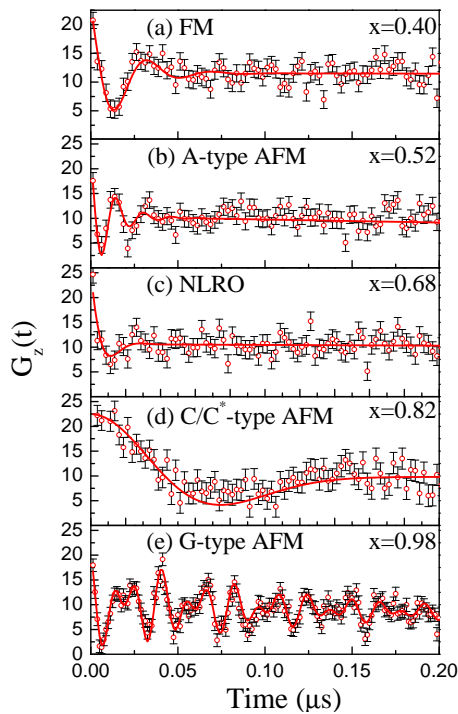}
   \caption{Muon spin relaxation data
at T=1.8~K for La$_{2-2x}$Sr$_{1+2x}$Mn$_2$O$_7$ with (a)
$x$=0.40 (FM), (b) $x$=0.52 (A-type AFM), (c) $x$=0.68 (NLRO), (d)
$x$=0.82 (C/C*-type AFM) and (e) $x$=0.98 (G-type). Two
oscillation frequencies can be observed for $x$=0.98 and a very
strongly damped oscillation for $x$=0.68. The solid lines are fits
to the data
 using  Eq.\ref{ch3:Gztotal}.} \label{ch3:musr_raw2k}
\end{figure}

Fig.~\ref{ch3:musr_raw2k} shows $\mu$SR spectra collected at
$T$=1.8~K (at PSI). Two
oscillations can be distinguished for the G-type AFM sample,
$x$=0.98, and only one damped oscillation for all others, although
for $x$=0.68 the damping is extremely large. The data were fitted
using  Eq.~\ref{ch3:Gztotal} and the obtained parameters are
listed in Table~\ref{ch3:frequency}.
\begin{table}[htbp]
\vspace{0.25cm} \caption{Parameters obtained by fitting the
$\mu$SR spectra for La$_{2-2x}$Sr$_{1+2x}$Mn$_2$O$_7$ at 1.8~K
(Fig.~\ref{ch3:musr_raw2k}). } \vspace{0.25cm}
\begin{tabular}{lcccc}
\hline \hline
Sample       & $\nu_{\mu}$ &$1/T_2$ & $1/T_1$ & $1/(2\pi \nu_{\mu}T_2)$  \\
  &  (MHz) &($\mu$s$^{-1}$) & ($\mu$s$^{-1})$& (\%)   \\
\hline
x=0.40       &  28(1)       & 57(6)   & 0.005(1) & 0.32(4)  \\
x=0.52       &  68(4)       & 110(14) & 0.10(1)  & 0.26(4)  \\
x=0.68       &  30(5)       & 150(23) & 0.06(2)  & 0.80(18)  \\
x=0.82       &  5.4(2) & 20(1) & 0.03(1)  &  0.58(15)     \\
x=0.98       &  47(2)/76(2) & 16(2)   & 0.065   & 0.054/0.034(0.6)  \\
 \hline
 \hline
 \end{tabular}
\label{ch3:frequency}
\end{table}
The source of the internal field is attributed to the aligned
manganese moments in the vicinity of the muon. As a function of
hole doping, $x$, the values of $\nu_{\mu}$ indicate changes in
the local field when the spin structure varies. The precession
frequency for $x$=0.40 is $\nu_{\mu}$=28(1)~MHz at $T=1.8$~K (which
corresponds to an internal field of $B_{\mu}$=0.21(1)~T),
similar to that found for the ferromagnetic system, $x$=0.3
\cite{Heffner1998}. The observation of two muon frequencies for
the G-type AFM compound, $x$=0.98, suggests that the muon occupies
magnetically inequivalent sites in this compound. In oxides the
muon site is usually
about 1~\AA~ from an oxygen atom
\cite{Brewer1990}. In the
layered manganite structure there are three inequivalent oxygen sites
O(1), O(2) and O(3) and one likely muon site is close to O(2) (0, 0, 0.2)
\cite{Heffner1998}.

The longitudinal relaxation rate, $1/T_1$, at $T=1.8$~K is 2--3 orders of
magnitude smaller than the transverse relaxation rate, $1/T_2$,
indicating that the relaxation of oscillations is dominated by a
{\it static} distribution of internal local fields determined by
the neighbouring Mn spins at the muon site \cite{ISIS}.
Nevertheless, some {\it dynamic} broadening, {\it i.e.}
fluctuations of Mn spins, is also present since $1/T_1$ is nonzero.
The relative width of the field distribution can be parameterized by
$1/(2\pi \nu_{\mu}T_2)$ (Table~\ref{ch3:frequency}) which is very
large ($\approx 0.80$) for the NLRO sample ($x$=0.68) and quite
small ($< 0.055$) for
 the G-type AFM compound  ($x$=0.98) that
 has only 2\% Mn$^{3+}$ ions mixed
with Mn$^{4+}$ ions. A large relative width
can be associated
with a large distribution of Mn$^{3+}$ and Mn$^{4+}$ ions
(resulting in a large
degree of local magnetic inhomogeneity),
or with the coexistence of different magnetic phases.

\begin{figure}[h]
\centering
 \includegraphics[width=6.3cm]{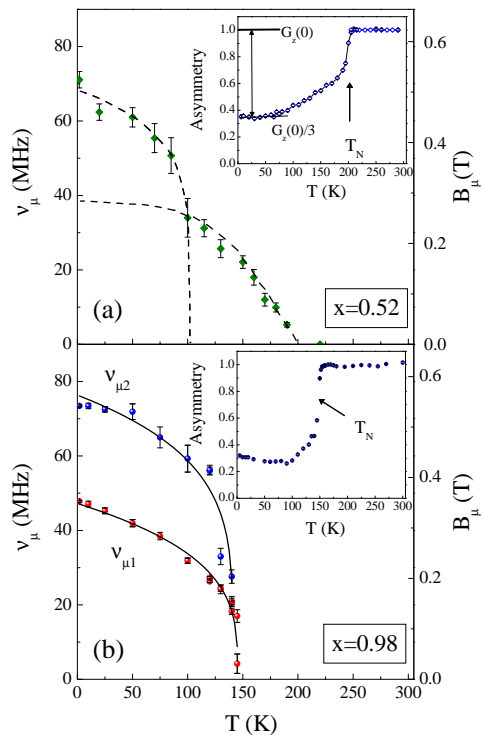}
 \caption{Temperature dependence of
    the zero-field muon precession frequency, $\nu_{\mu}$, and internal field, $B_{\mu}$,
       for (a)  $x$=0.52 and (b) $x$=0.98. The inset shows the temperature
   dependence of the relaxing asymmetry, A$_{\rm rlx}$ for the two compositions.}
    \label{ch3:x52_x98freq}
\end{figure}

The temperature dependence of $\nu_{\mu}$(T), (and
hence $B_{\mu}$(T)) for $x$=0.52 and $x$=0.98 is shown in
Fig.~\ref{ch3:x52_x98freq}. For $x$=0.52 we find a discontinuity
in $\nu_{\mu}$(T) around 100~K. This arises because of the
presence of two magnetic phases, A-type AFM and a CO phase similar
to that found in $x$=0.5 \cite{Argyriou2000,Chatterji2000}, which compete with
each other in the region between 100~K and 200~K. Below 100~K
the internal field increases significantly
due to an additional magnetic transition.
This transition coincides with a broad maximum
in $1/T_1$ \cite{DPHIL2001}, signifying a freezing of the local spins.
This effect is strikingly similar to that observed
in a bilayer cuprate \cite{Niedermayer1998},
in which a {\it freezing} of spins of doped holes
is superimposed on AFM long-range ordered
Cu$^{2+}$ spins.
 The
slow variation of the relaxing asymmetry, A$_{\rm rlx}$ (inset of
Fig.~\ref{ch3:x52_x98freq}) between $G_z$(0) (at 200~K) and $G_z$(0)/3
(below 100~K) further confirms the {\it two-phase model} whose volume
fraction is temperature dependent and that the whole
sample is magnetic below 100~K.
The behavior of the $x$=0.52
compound is in contrast to the G-type AFM compound, $x$=0.98, for
which the drop in the initial
asymmetry is much sharper (see inset Fig. 2(b)) and
both precession frequencies $\nu_{\mu1}(T)$ and
$\nu_{\mu2}(T)$ can be fitted to $\nu_{\mu i}(T)=\nu^0_{\mu i}
(1-T/T_{\rm N})^{\beta}$ (Fig. 2(b)).
\begin{figure}[htbp]
\centering
 \includegraphics[width=6.5cm]{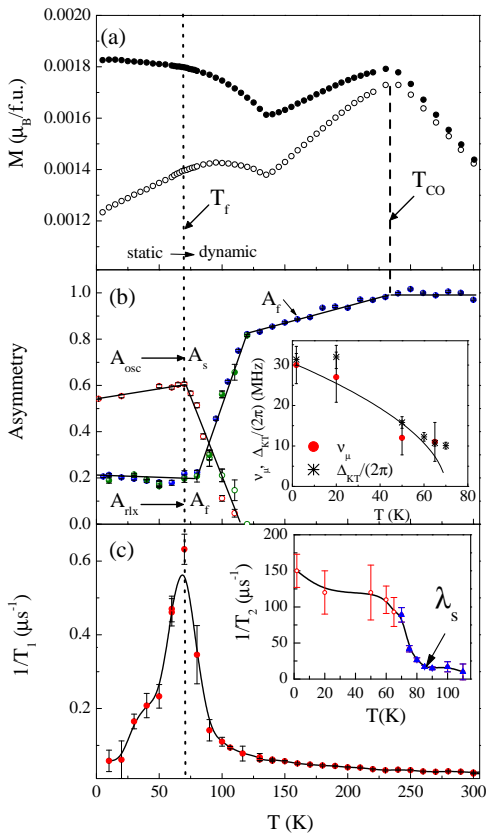}
    \caption{Temperature dependence  of
   different fitting parameters for $x$=0.68;
     (a) the zero-field cooled (open circles) and field-cooled (closed circles)
     magnetization in 10~mT;
   (b) the static amplitude ($A_{\rm osc}$ for $T<T_{\rm f}$
   and $A_s$ for $T>T_{\rm f}$) and
   the dynamic amplitude
   ($A_{\rm rlx}$ for  $T<T_{\rm f}$
   and $A_{\rm f}$ for $T>T_{\rm f}$); inset shows
    $\nu_{\mu}(T)$ and $\Delta_{\rm KT}/(2\pi)$;
   (c) longitudinal relaxation rate, 1/$T_1$,
    (the inset shows the transverse relaxation rate, 1/$T_2$,
    and slow component, $\lambda_{\rm s}$).
  The freezing temperature $T_{\rm f}$ (vertical dotted line)
  separates the static and the dynamic regions. The vertical dashed
  line shows $T_{\rm CO}$ defined as the maximum in
the magnetization data, in agreement with electron diffraction studies
\cite{Li2001b}.
  The solid lines are a guide to the eye.}
   \label{ch3:x68_musr}
\end{figure}
This procedure, although usually applicable only
to the asymptotic critical regime,
gives
a reasonable description of the data over the entire temperature
range (see Fig.~\ref{ch3:x52_x98freq}(b)),
yielding $T_{\rm N}$=145.0(5)~K in agreement
with neutron powder diffraction \cite{Ling2000}
and $\beta$=0.24(2)
(close to the value  corresponding
to a 2D XY system ($\beta$=0.23) \cite{Bramwell1993,Rosenkranz2002}).

One of the most intriguing regions of the Mn$^{4+}$-rich phase
diagram is $x$=0.66--0.74 where NLRO has been detected
using neutron diffraction \cite{Ling2000}.
Data for $x$=0.68 at $T$=1.8~K shown in Fig.~1(c)
were analyzed using a relaxation function
which was
(1) Eq.~1 for $T\leq70$~K, (2)  $G_z(t)=A_{\rm s}{\rm
e}^{-\lambda_{\rm s}t} +A_{\rm f}{\rm e}^{-\lambda_{\rm f}t}$ for
$70<T \leq 120$~K and (3) $G_z(t)=A_{\rm f} {\rm e}^{-\lambda_{\rm f} t}$
for $T>120$~K and the results of the fitting are presented in
Fig.~\ref{ch3:x68_musr}. At low temperatures, the strongly damped
oscillation frequency decreases upon warming from $\sim 30$~MHz at
2~K to $\sim 10$~MHz around 65~K (inset to
Fig.~\ref{ch3:x68_musr}(b)), suggesting the development of a
quasi-static field at the muon-site indicative of short-range
order. In the low temperature regime we found that modelling data
using a Gaussian Kubo-Toyabe relaxation function
parameterized by $\Delta_{\rm KT}/(2\pi)$
gave similar results (see inset of Fig.~\ref{ch3:x68_musr}(b)).

The broad static distribution of local fields
is quantified by the large $1/T_2$
values  which is replaced by the slow dynamic
relaxation rate $\lambda_{\rm s}$ above 70~K (inset to Fig.~3(c)).
The damped oscillations disappear above 70~K, the temperature at which
$1/T_1$ exhibits a pronounced maximum, corresponding to
the slowing-down of spin fluctuations, and we identify this
temperature as the {\it spin-freezing} temperature, $T_{\rm f}$.
The amplitudes of the two components $A_{\rm osc}$ and $A_{\rm
rlx}$ shown in Fig.~3(b) provide information about the volume
fractions of the sample where muons experience static or dynamic
spin-lattice relaxation, respectively.
Above $T_{\rm f}$ the
oscillatory component, $A_{\rm osc}$, (and the corresponding
frequency $\nu_{\mu}$) disappears and is
replaced by a slowly fluctuating fraction with amplitude $A_{\rm s}$
(Fig.~\ref{ch3:x68_musr}(b)).
This fraction decreases
upon heating above $T_{\rm f}$ and disappears around 120~K.
At high temperatures the main source of depolarization is the
fast fluctuating spins and the corresponding volume fraction,
$A_{\rm f}$, increases linearly on warming (for 120~K$<T<T_{\rm CO}$).
This suggests that in this temperature range
the short-range (CE-type) or dynamic CO dominates the muon depolarization,
as found for $x=0.5$ \cite{Andreica2000} and La$_{1-x}$Ca$_x$MnO$_3$ \cite{Kim2000}.
The zero-field-cooled magnetization decreases almost
linearly with temperature for 120~K$<T<T_{\rm CO}$
due to the formation of short-range antiferromagnetic correlations
in the CO region
(Fig.~\ref{ch3:x68_musr}(a)).

In the limit of fast fluctuations
one can estimate the correlation time, $\tau$, of magnetic fluctuations
by assuming that the static width of
the field distribution
is of the order of  the average magnetic
field sensed by muons at the lowest temperature
($\gamma_{\mu}\Delta B_{\mu} \sim 2 \pi \nu_{\mu}$).
We estimate
short correlation times  associated with fast fluctuations
 in the range $\tau_{\rm f}$=10$^{-12}$--10$^{-13}$~s
 for all samples, in good agreement with other values
reported for bilayer manganites
\cite{Perring1997,Feng2001}.
For $x$=0.68, the additional slowly fluctuating
component observed between 70~K and 120~K
fluctuates with a correlation time
$\tau_{\rm s}$=10$^{-9}$--10$^{-11}$~s.
The existence of
{\it slow} and {\it fast} dynamics between $T_{\rm f}<T<120$~K
for $x$=0.68 further sustains the {\it phase separation} scenario
in which charge-ordered and ferromagnetic (or A-type or C-type AFM)
regions may coexist in the NLRO region,
as found in La$_{1-x}$Ca$_x$MnO$_3$ \cite{Heffner2001}.

\begin{figure}[htbp]
\centering
  \includegraphics[width=6.35cm]{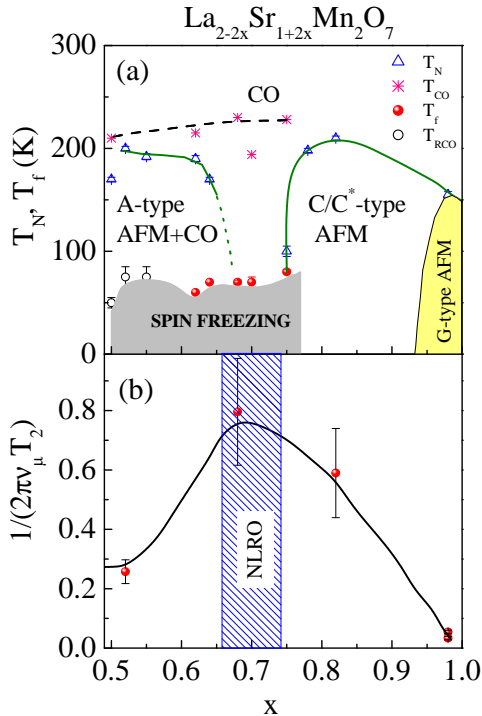}
     \caption{(a) Phase diagram
    of La$_{2-2x}$Sr$_{1+2x}$Mn$_2$O$_7$ ($0.5<x<1$).
The charge ordering
temperatures were obtained from magnetization data \cite{DPHIL2001}.
Solid and dashed lines are guides to the eye.
   (b) Degree of magnetic inhomogeneity
at 1.8~K expressed by $1/(2\pi \nu_{\mu}T_2)$.}
   \label{ch3:correlation_time}
\end{figure}
The freezing mechanism
begins with the formation of magnetic clusters at high temperatures.
Thermal disorder opposes this tendency, so that with decreasing
temperature the regions of correlated spins become larger and the
spin system is subdivided into independent magnetic clusters of
different size with a distribution of blocking temperatures.
If a cluster is blocked
for a time greater than the time window of the  $\mu$SR technique,
it will appear {\it frozen} or {\it quasistatic} and hence $A_{\rm rlx}$
decrease reflects a progressive
blocking of clusters by cooling down towards $T_{\rm f}$,
where all clusters are blocked.
Spin-freezing phenomena
are found
in numerous systems, ranging from
 spin-glasses, such as Cu-Mn alloys \cite{Hunt1999}
 to lightly-doped  cuprates,  La$_{2-x}$Sr$_x$CuO$_4$ \cite{Niedermayer1998}.
 In our layered manganites
we find that
short-range charge-ordered fluctuations are {\it dynamic} at
high temperatures ($T<T_{\rm CO}$)
and become {\it static} below $T_{\rm f}$
(or in the reentrant charge-ordered region).
We note that there has been evidence for spin freezing in manganese
perovskites ascribed to magnetic cluster formation and/or
magnetic polarons \cite{Papavassiliou2001,Allodi2001}
or stripe-like structures \cite{Hotta2001}.

In conclusion, our $\mu$SR study
of electron-doped bilayer manganites
yields {\it local} magnetic information
and provides evidence for
magnetic order, phase separation
and {\it spin freezing}.
The results are summarized in the phase diagram shown
in Fig.~4(a).
Spin freezing is observed
as a maximum in $1/T_{1}$ for
samples with $x$=0.52-0.75 \cite{DPHIL2001}.
The spin frozen state thus exists
in the region
of the phase diagram with LRO ($x$=0.52-0.62) together
with a LRO state (AFM for $x$=0.52)
and also in the region for which neutron diffraction
failed to detect any LRO.
We note that the {\it spin-frozen} region
of the phase diagram overlaps closely with
a region in which {\it reentrant charge-ordering}
has been suggested to exist
\cite{Chatterji2000,Li2001b,Yuan1999,Dho2001}.
We can quantify the degree of
local magnetic inhomogeneity at low temperatures
using $1/(2\pi \nu_{\mu}T_2)$ (Fig.~4(b)) which
is largest for the $x=0.68$ composition in the NLRO region
\cite{Ling2000}.

We thank Ishbel Marshall, Anke Husmann, Alex Amato and Stephen Cottrell
for help during experiments and EPSRC for financial support.
AIC thanks ORS and University of Oxford for financial
support. This work was sponsored in part by
the US Department of Energy Office of
Science under Contract No. W-31-109-ENG-38.

\end{document}